\documentclass[conference, letterpaper]{IEEEtran}
\IEEEoverridecommandlockouts
\usepackage{cite}
\usepackage{amsmath,amssymb,amsfonts}
\usepackage{algorithmic}
\usepackage{graphicx}
\usepackage{textcomp}
\usepackage{geometry}
\usepackage{xcolor}
\def\BibTeX{{\rm B\kern-.05em{\sc i\kern-.025em b}\kern-.08em
    T\kern-.1667em\lower.7ex\hbox{E}\kern-.125emX}}
\begin{document}

\newgeometry{top=1in, left=0.75in, right=0.75in, bottom=0.75in}

\title{Ontology-driven Reinforcement Learning for Personalized Student Support\\
\thanks{This work was supported in part by the National Science Foundation under Grants 1913809 and 2121277.}
}

\author{\IEEEauthorblockN{Ryan Hare}
\IEEEauthorblockA{\textit{Dept. of Electrical and Computer Engineering} \\
\textit{Rowan University}\\
Glassboro, NJ \\
harer6@rowan.edu}
\and
\IEEEauthorblockN{Ying Tang}
\IEEEauthorblockA{\textit{Dept. of Electrical and Computer Engineering} \\
\textit{Rowan University}\\
Glassboro, NJ \\
tang@rowan.edu}
}

\maketitle

\begin{abstract}
In the search for more effective education, there is a widespread effort to develop better approaches to personalize student education. Unassisted, educators often do not have time or resources to personally support every student in a given classroom. Motivated by this issue, and by recent advancements in artificial intelligence, this paper presents a general-purpose framework for personalized student support, applicable to any virtual educational system such as a serious game or an intelligent tutoring system. To fit any educational situation, we apply ontologies for their semantic organization, combining them with data collection considerations and multi-agent reinforcement learning. The result is a modular system that can be adapted to any virtual educational software to provide useful personalized assistance to students.
\end{abstract}


\section{Introduction}
In the past several years, there has been an exponential increase in developments in artificial intelligence (AI), and in many respects, these new developments still fall short of the potential or creativity of human intelligence. However, AI does have great potential to augment human intelligence and to assist in a variety of tasks \cite{aiaugmentation}. AI assistance is especially powerful and notable in areas that cannot otherwise be improved due to lack of human resources or overall resource limitations. In this paper, we focus on one such area; personalized education. To be specific, we propose a general-purpose framework leveraging ontologies as a knowledge model and multi-agent reinforcement learning as an AI "brain" to create a modular and widely-applicable solution to the task of personalized student support. Our theoretical method, as written, can apply to a variety of virtual educational systems and software with the ultimate goal of providing students with timely and personalized feedback.

The challenge of personalized learning has been ongoing since at least 2008, when the National Academy of Engineering declared the advancement of personalized learning one of its 14 grand challenges in engineering \cite{grandchallenges}. And personalized learning does prove a challenging problem \cite{challengeofpersonalization}; due to instructor resource constraints, limited school budgets, and large classroom sizes \cite{classsizes}, it is infeasible for human intelligence to individually tutor every single student in a way that fits their ideal preferences. Thus, typical educational methods lean more toward a one-size-fits-all approach; that is, a teaching method that is useful and effective for all students. However, there are always outliers, and one-size-fits-all methods still tend to leave students behind, especially in situations where those students have a shaky background or a fundamental misunderstanding \cite{onesizeissues}. And ultimately, falling behind in lessons often leads many students to drop out from classes, suffer poor grades, fall behind on assignments, or resort to time-consuming self-exploration.

Fitting with AI's ability to augment human intelligence, AI is easily integrated into virtual educational systems such as serious games and intelligent tutoring systems \cite{aiineducation, aiineducation2}. With AI and virtual educational systems, students can be given active, timely feedback, regardless of whether they are learning in the classroom, at an after-school study group, or at home. Furthermore, the AI can actively prompt and engage with students, helping them to overcome situations where they may fail to even formulate a proper question \cite{aiengagement}. Finally, AI can easily be integrated with a lesson plan or educational database to answer questions or address any misunderstandings that a student may have without the need for instructor intervention. With AI as a study aid, well-performing students can cover their areas of difficulty while poor-performing students can catch up without excessive instructor intervention.

AI integration in virtual educational systems first require some knowledge database or lesson plan to guide education, track students' progress, and allow for human input to ensure accurate lessons. For our proposed method, this knowledge database is created with an ontology; a type of data structure used to model and organize knowledge \cite{whatontology}. With their structure, ontologies are a powerful framework, usable for organizing educational content in a hierarchical manner while creating a format that can be easily connected with an AI. In literature, ontologies have shown effectiveness for student modelling for recommender systems \cite{ontomodeling} and intelligent tut-

\newgeometry{top=0.75in, left=0.75in, right=0.75in, bottom=0.75in}

\noindent oring systems \cite{ontomodellingits}, simplifying lesson planning while accounting for educational standards \cite{ontostandards}, and for representing general educational content \cite{ontocontent}.

\begin{figure*}[!ht]
  \centering
  \includegraphics[width=5in]{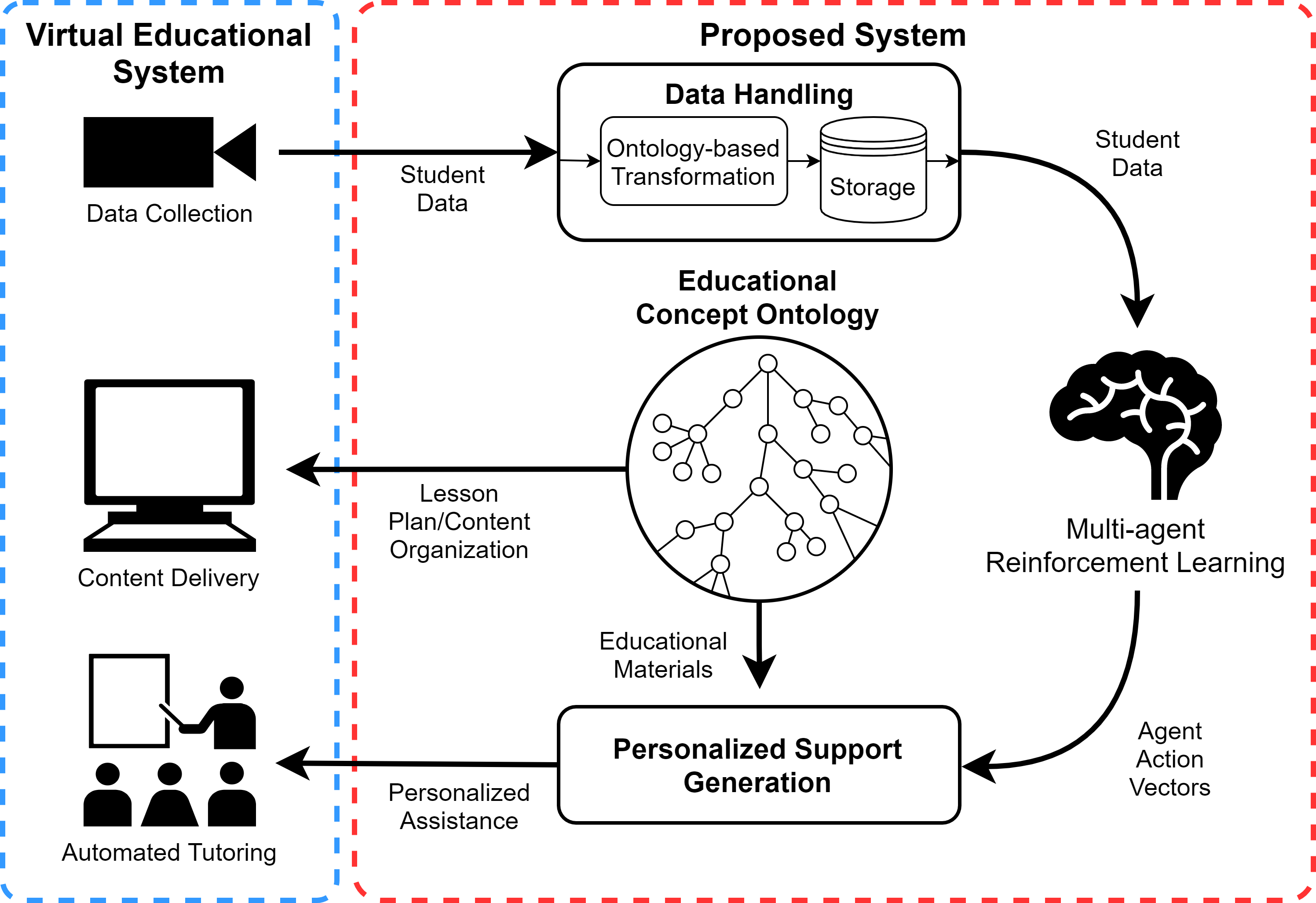}
  \caption{Full diagram of the proposed system showing all four components and connections to a virtual educational system.}
  \label{fig:system}
\end{figure*}

With an ontology representing the domain knowledge that a virtual educational system intends to teach, the remaining components of the proposed system serve as the "brain" of the AI. We adapt multi-agent reinforcement learning (RL) \cite{reinforcementlearning} to act as an adaptive AI tutor. This dynamic adaptability both guarantees that the AI will eventually perform optimally \cite{rlconvergence}, and allows the AI system to account for shifting trends in student behavior. In our prior work, we developed and tested a multi-agent RL approach using experience sharing to improve agent training performance \cite{priormulti}. We verified the positive performance impact of this method \cite{priortoe}, and have successfully applied this method within an educational serious game, showing a positive impact on student performance \cite{priortoe}.

However, our prior work applied the multi-agent reinforcement learning to a single case, dealing with a small subset of topics within a single domain. It wasn't straightforward to add new topics to the system, and furthermore, the system wasn't generalized for wider applications. With the proposed work, we aim to extend the system to support multiple connected RL agents using the ontology structure for organization. By adapting our prior experience sharing method, newly-added agents can boost their performance using shared experience, giving them a jump-start at the beginning of training to avoid a period of poor performance at the start. The result is a modular, easily-extensible student support system that can function in various virtual educational systems on various topics.


\section{Proposed System Architecture} \label{sec:pro}
For the system detailed in this paper, we first provide an overview of the various components before detailing each component part. As shown in Figure \ref{fig:system}, the proposed system has 4 main components:

\begin{enumerate}
    \item Educational Concept Ontology: The ontology is the core of any integrated system, providing a structured and hierarchical representation of the relevant domain knowledge. The structure and organization of the ontology informs all other components.
    \item Data Collection, Transformation, and Storage: With the ontology storing a structured representation of relevant domain knowledge, any integrated system must also collect, transform, and store data on student performance. As the data collected are not necessarily in the required format for the system, converting the data to the correct set of variables is necessary. Overall, data are necessary to inform the system's "brain", the reinforcement learning.
    \item Reinforcement Learning: The AI "brain" of the system, the reinforcement learning agents are responsible for adjusting system behavior and deciding what feedback students receive. The output of the reinforcement learning agents informs the personalized assistance generation module.
    \item Personalized Assistance Generation: The personalized assistance generation module pulls domain knowledge from the ontology and, using inputs given by the "brain", constructs the personalized assistance that students receive when interacting with an integrated system.
\end{enumerate}

The following sections discuss these components in greater detail.

\subsection{Ontologies for Education} \label{sec:ont}
\begin{figure}[!h]
  \centering
  \includegraphics[width=\columnwidth]{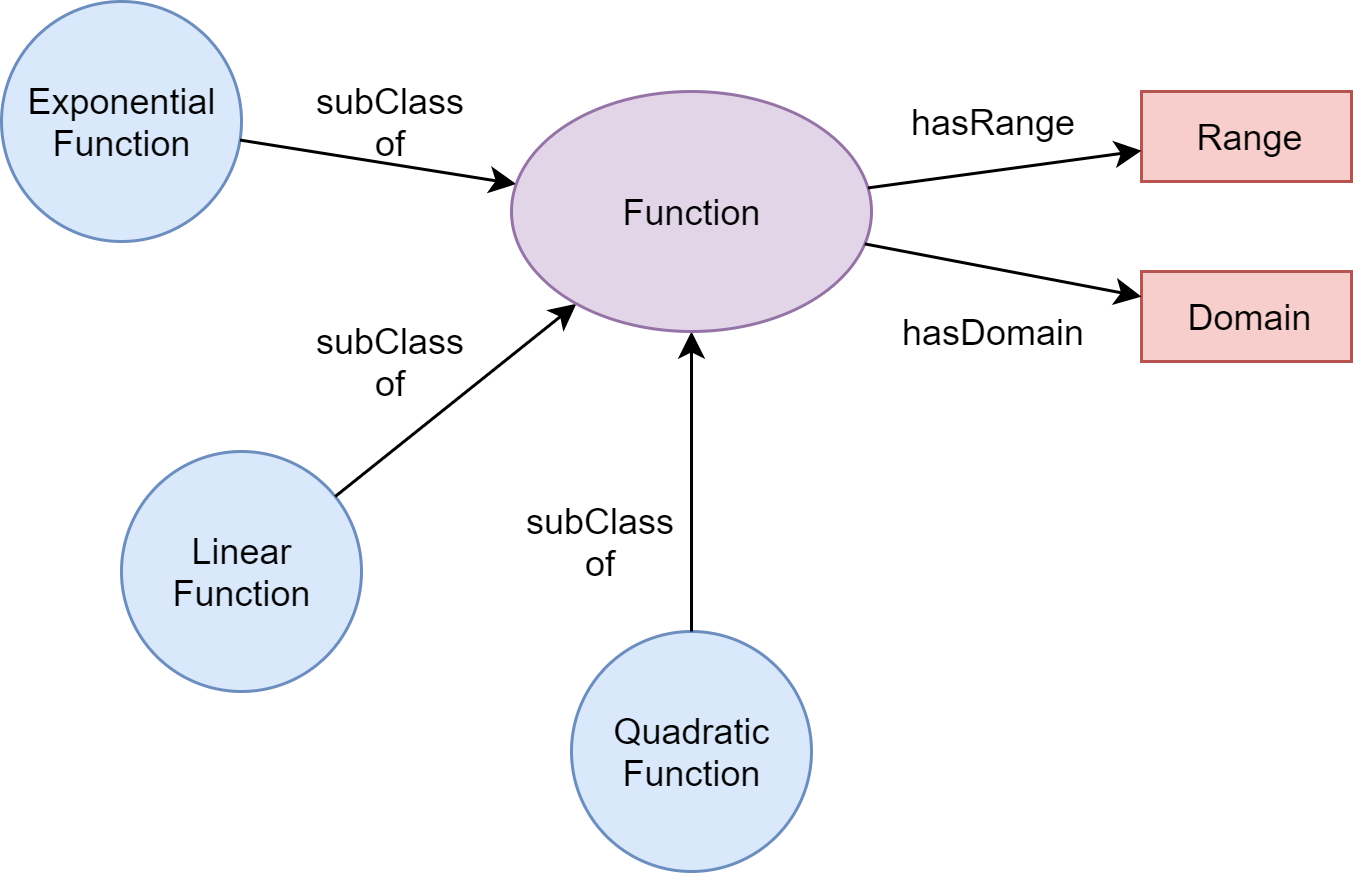}
  \caption{A simplified example ontology that focuses around mathematical functions, showing a few possible sub-concepts and how they're related.}
  \label{fig:onto}
\end{figure}

The first component of the proposed system is an educational ontology. As was previously mentioned, ontologies have seen application in education for student modelling, grading, and the organization of educational material. For the proposed system, the ontology within provides a structured representation of the domain knowledge that a virtual education system aims to teach. An example ontology dealing with mathematical functions is provided in Figure \ref{fig:onto}.

For this application, ontologies were chosen due to their past successes in educational applications, including student modelling \cite{ontomodeling, ontomodellingits}, and educational content organization \cite{ontocontent, ontostandards}. Furthermore, the semantic connections used allow for a more meaningful structure, informing the lesson plan and educational content presentation of any connected system. In other words, topics can be organized similarly to a lesson plan, helping to inform both AI assistance and the overall order of content presentation.

The ontology determines the order of operations that the system follows when teaching a student, as well as the topics that are assigned reinforcement learning agents. As the RL agents are meant to specialize in topics, agent-augmented nodes in the ontology must be granular enough where a specialized agent can provide relevant and useful assistance, while still being broad enough to not require a massive number of RL agents. In the case of Figure \ref{fig:onto}, we might assign an RL agent to the sub-classes "linear function", "quadratic function", and "exponential function", as well as one for the properties of domain and range. Then, any topic under the selected topics can be used to inform tutoring and generate personalized recommendations, as will be discussed in a later section.

Formally, an ontology for this application can be modelled as a directed acyclic graph $G = \langle C, E, \rho, D, \alpha \rangle$ such that:
\begin{itemize}
    \item $C = \{c_i\}$ is the set of concepts contained within a given ontology, visualized as nodes in the graph. Each concept $c_i$ represents a specific piece of knowledge or topic within the ontology's target domain.
    \item $E = \{c_i, c_j, p_k\}$ is a set of directional edges that connect a topic $c_i \in C$ to another topic $c_j \in C$. The semantic relationship between these topics is defined as a property $p_k$. As edges are directed, concepts should be connected in a hierarchical manner.
    \item $\rho(c_i) = \{\varrho_j\} \forall c_i \in C$ is a set of attributes or properties $\varrho_j$ associated with each concept $c_i \in C$ to further define each concept. Properties depend on application, but could represent difficulty level, relevance to an overarching educational objective, or any other useful properties to assist system configuration.
    \item $D$ is a database of educational materials such that for each concept $c_i \in C$ there exists a subset $D_{c_i} \in D$ that contains educational materials related to concept $c_i$. Educational materials are used to tutor students, and could be anything from simple text tutorials to videos, example problems, or interactive simulations, for example.
    \item $\alpha$ is a set of reinforcement learning (RL) agents assigned to select concepts in $C$. For some concepts $\{c_i\} \subseteq C$ there exists an RL agent $\alpha_{c_i}$ assigned to that concept.
\end{itemize}

\subsection{Data Collection and Storage} \label{sec:dat}
With the ontology defined, the second step in system creation is student data collection. Given an ontology $O^*$ and its set of concepts $C^*$, the system collects student data for each concept $c_i \in C^*$. By matching student data to each topic in the ontology, the system has a solid understanding of a student's performance in each of the topics. We formalize this in the proposed system by stating that each concept in the ontology stores a student data vector, $\mathbf{x_{c_i}}$, such that this vector contains a set of generic performance metrics that can indicate a student's performance in any given topic.

The specific performance metrics used depend on implementation, but should be general-purpose such that any topic can use the same metrics. This shared form for every node is important to the reinforcement learning, as will be discussed in Section \ref{sec:rl}. To give some examples, $\mathbf{x_{c_i}}$ could be defined as $\mathbf{x_{c_i}} = \begin{pmatrix} a & b & c \end{pmatrix}$, where $a$ is a measure of competency, such as a quiz score, $b$ is a measure of timing, such as a time-to-completion measurement, and $c$ is a numerical indicator of engagement, such as number of interactions made or time spent idle. Ultimately, the data points chosen depend on implementation.

\subsection{Data Transformation} \label{sec:trans}

As mentioned, each data vector $\mathbf{x_{c_i}}$ has a shared form for every topic, and is meant to represent a general-purpose vector (referred to here as $\mathbf{x}$) of student performance metrics that can be used for any educational topic or scenario. However, the inherent diversity in data collection methods also requires the proposed system to account for variation in collected student data. Therefore, it is necessary for the proposed system to transform incoming data from some unknown set of collected metrics to a known list of accepted metrics.

\begin{figure}[!h]
  \centering
  \includegraphics[width=\columnwidth]{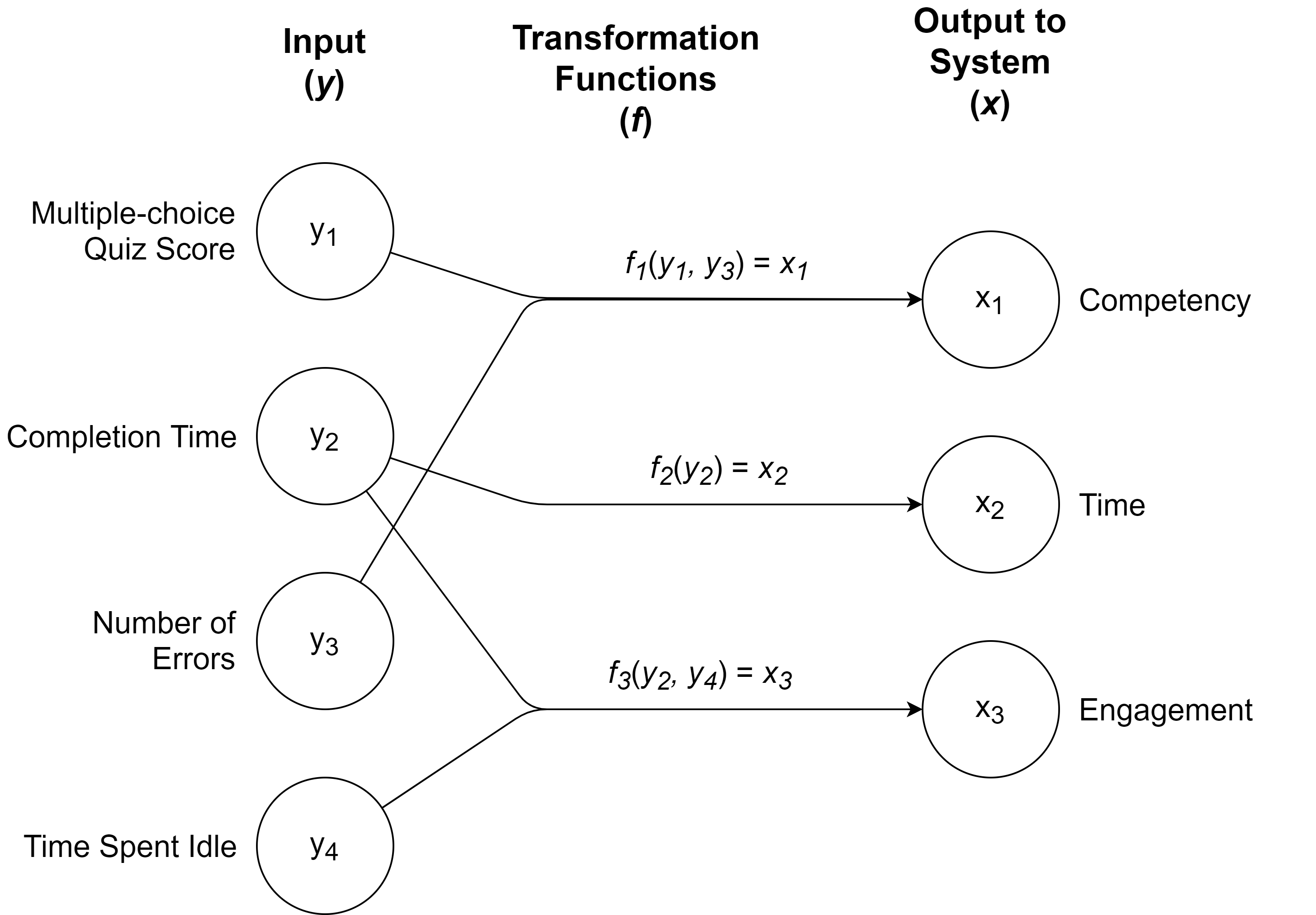}
  \caption{An example transformation graph showing how some unknown metrics from $y$ with varied descriptors are mapped to $x$, with appropriate transformation functions assigned along the connecting edges.}
  \label{fig:transdata}
\end{figure}

Connecting with other sections of the proposed system architecture, data transformation begins with a semantic representation of the target data metrics in $\mathbf{x}$. With this representation, it is necessary to assign a set of descriptors to each metric in $\mathbf{x}$. Then, if the system has a new vector $\mathbf{y_{c_i}}$ of input data with format $\mathbf{y}$, containing some varied set of metrics each also with descriptors attached, $\mathbf{y}$ can be connected within the data transformation ontology to $\mathbf{x}$. With connections formed, let $f : \mathbf{y} \xrightarrow{} \mathbf{x}$ define a set of transformation functions that map each element in $\mathbf{y}$ to an element in $\mathbf{x}$, performing necessary data transformation and normalization to translate $\mathbf{y}$ into $\mathbf{x}$. Figure \ref{fig:transdata} shows an example of how this data transformation can be formalized. Section \ref{sec:ex} shows a more in-depth example of this data transformation principle.

\subsection{Multi-agent Reinforcement Learning} \label{sec:rl}
Reinforcement learning \cite{reinforcementlearning} is a machine learning paradigm wherein an intelligent agent interacts with some environment.  As it interacts it makes numerical observations of the current environment, which are called states. Based on the observed state, the agent selects one or more actions from a set of possible actions, with those actions, in turn, having some effect on the state. The agent is then given a numerical reward, which is determined by the state change, the action chosen, or even some external factor. The agent observes the reward and the new environmental state, and repeats the process. By applying this paradigm as our "brain", our proposed system has a set of these reinforcement learning agents through multi-agent reinforcement learning with experience sharing, a novel approach that we've proven in prior work on the same task of personalized student assistance \cite{priormulti}. Each agent in the system is then assigned to a specific topic in the ontology.

Reinforcement learning provides a natural fit for the problem of personalized student assistance. Given the established data vector, $\mathbf{x_{c_i}}$, that stores student data relative to a given topic $c_i \in C^*$ in our example ontology $O^*$, this data can represent an environmental state for an RL agent. The iterative online learning that reinforcement learning agents provide means that system developers do not have to encode complex decision logic to provide personalized support, instead relying on the RL agents to naturally learn optimal behavior. Furthermore, since reinforcement learning is an ongoing, online process, shifting trends in student preferences or slight changes in lesson planning can be adapted to, ensuring that students always receive appropriate personalized assistance.

Formally, the reinforcement learning problem is presented as a multi-agent Markov decision process (MDP) such that each agent in our ontology $O^*$, labelled as $\alpha_{c_i} \in \alpha^*$ has an $MDP = \langle S, A, R, p, \gamma \rangle$:

\begin{itemize}
    \item $S$ is a given agent's finite state space. In this application, a state observation at a given time step is an observation of $\mathbf{x_{c_i}}$ at time step $t$, also defined as $s_t \in S$. Thus, the total size of the state space is defined by the dimensionality and variable range of $\mathbf{x_{c_i}}$.
    \item $A$ is a given agent's finite action space. For this application, agent actions are defined as a vector, $\mathbf{a} \in A$ such that each element of $\mathbf{a}$ represents a property of personalized student assistance. The action chosen at a given time step is defined as $\mathbf{a}_t \in A$ In this case, the agent does not select a single action, but rather, assigns a value to every property in the action vector. 
    
    Section \ref{sec:ex} gives an in-depth example of how an action vector might be defined using properties of personalized student assistance.
    \item $R$ is the reward function defined as $R(s, a, s') \xrightarrow{} \mathbb{R} \forall s, s' \in S, a \in A$. Reward is dependent on application, but an ideal reward function will encourage certain behavior. For example, defining the reward function to return a positive reward for improvements in student performance is a natural configuration for the proposed system.
    \item $P$ is the transition probability dynamics of the system, defined as $P(s' | s, a) \xrightarrow{} [0, 1] \forall s, s' \in S, a \in A$. It determines the probability of arriving in some new state based on the prior observation and the action chosen. This function is challenging to predict in a human-centric environment like the proposed system.
    \item $\gamma$ is the discount factor such that $\gamma \in [0, 1]$. The discount factor applies a weight to future rewards. A value of 0 makes the agent consider only the immediate first step in its decision-making, while a value of 1 makes the agent consider all possible future rewards with no weight applied.
\end{itemize}

Given the possibly large dimensionality of the state space and the continuous output format for agent actions, choosing an appropriate reinforcement learning method is also important for successful implementation. Deep reinforcement learning methods are useful for handling large, non-discrete state spaces, and the method adapted here made use of Deep Q-learning as the central reinforcement learning algorithm. However, given the continuous nature of the action vector, a certain subset of RL algorithms, deep policy gradient methods fit this application better. Unlike other RL methods which rely on estimating the outcome of discrete actions, deep policy gradient methods can work with continuous actions. As such, they are typically used in robotics, but are highly applicable in the proposed system. Recently-developed methods like deep deterministic policy gradient (DDPG) \cite{ddpg}, twin delayed DDPG \cite{td3}, soft actor-critic \cite{sac}, or proximal policy optimization \cite{ppo} all fit this application, and would still work with our multi-agent experience sharing method.

\subsection{Generating Personalized Assistance} \label{sec:ass}
Finally, given an action vector, $\mathbf{a^*}$, from the RL agent working on concept $c_i$, the personalized assistance generation module is responsible for translating both the action vector and the database content subset $D_{c_i}$ into tutoring material or assistance that can be provided to the student. Given the wide range of possible situations the proposed system could be applied to, the outcome of the personalized assistance generation module could translate these inputs into some interface, pop-up, change in a virtual environment, or other adjustment in order to appropriately deliver the target content to the student.

Given the wide range of possible virtual environments that the proposed system could be applied to, the specific form of this module will vary. A serious game might prefer a non-player character to approach the player, offering varied dialogue based on the given inputs. An intelligent tutoring system, meanwhile, might offer a more direct pop-up prompt containing a set of recommended study materials, practice problems, and tutorial videos. This module is the most flexible, and could even take advantage of new technologies in dynamic content generation such as large language models \cite{llmedu} or other recent methods in AI-generated content \cite{aigcedu}. On the other hand, a more simplified approach might assign properties to the database entries, enabling or disabling certain pieces of content based on $\mathbf{a^*}$. Rule-based systems could also apply \cite{rulebasededu}, only pulling certain content from the database determined by a set of expert-defined rules based on $\mathbf{a^*}$. Ultimately, there are many approaches to translate these two inputs into personalized student assistance, and different integrated systems will have a different "best" solution.

\section{Example Implementation \label{sec:ex}}
Given the system definition, we now provide a brief example of how such a system might be implemented within some educational software. In this example educational software, we can assume that the system provides educational material to students and intermittently administers quizzes to gauge each student's content knowledge. With that in mind, we first must consider the data collected by such a system, defining the data input vector, $\mathbf{y} = \{y_1, y_2, y_3, y_4, y_5\}$ with the following values:

\begin{itemize}
    \item $y_1$: A given student's percentage of correct answers on a quiz in range $[0, 1]$
    \item $y_2$: The total amount of time spent on a given section, in seconds
    \item $y_3$: The amount of time spent engaging with educational content (ex. reading material, answering questions), in seconds
    \item $y_4$: The total number of button presses or other interactions the student has made within a given section
    \item $y_5$: An emotional state estimated from webcam images of the student as they play, with $0$ indicating negative emotions, $1$ indicating neutral, and $2$ indicating positive
\end{itemize}

With those values defined, we can now determine the mapping functions to translate the given data into the system's input vector, $\mathbf{x} = \{x_1, x_2, x_3\}$, where:

\begin{itemize}
    \item $x_1$: A numerical score in range $[0, 1]$ indicating a student's competency on the provided content
    \item $x_2$: A numerical score in range $[0, 1]$ indicating a student's engagement with the provided content
    \item $x_3$: A numerical score indicating a student's current emotional state in range $[0, 1]$, where $0$ indicates that frustration might be present, while $1$ indicates a positive emotional state with no measured issues 
\end{itemize}

And then the mapping functions can determine the relationship between the actual gathered data and the data vector sent in to the RL system. Quiz scores ($y_1$) can easily map directly into the value for competence ($x_1$), as shown in Equation \ref{eq:2}. Engagement ($x_2$) can be determined based on the student's number of interactions with the system ($y_4$) and how they compare to some maximum value ($y_4^{max}$) determined through testing. As shown in Equation \ref{eq:3}, engagement can also take into account the percentage of time a student spends on educational content ($y_3$) compared to their total time in the game ($y_2$) to provide further insight into their engagement with the provided material. Finally, as shown in Equation \ref{eq:4}, emotional state can use a scaled version of the emotional measurements ($y_5$), but can be adjusted if the user is spending an above-average amount of time (with $y_2^{avg}$ indicating average time) on a given section.

\begin{equation}
\label{eq:2}
x_1 = y_1
\end{equation}

\begin{equation}
\label{eq:3}
x_2 = (\frac{y_4}{y_4^{max}} + \frac{y_3}{y_2})/2
\end{equation}

\begin{equation}
\label{eq:4}
x_3 = \begin{cases}
    \frac{y_5}{2} & y_2 \leq y_2^{avg} \\
    \max{(\frac{y_5}{2} - \frac{y_2 - y_2^{avg}}{y_2^{avg}}, 0)} & y_2 > y_2^{avg}
\end{cases}
\end{equation}

Given these mapping functions, we aim to demonstrate that the proposed system can be flexible, with a variety of acceptable data collection methods and data inputs all mapping to a standard format so that different systems can use similar RL architectures. An actual implementation could extend this basic system to include significantly more complex mapping functions, such as neural networks, machine learning methods, or statistical analyses. With $\mathbf{x}$ fully mapped, the system can use it as the RL state to generate the action vector. In our example system, we define the action vector $\mathbf{a^*} = \{a_1, a_2, a_3, a_4, a_5\}$, where:

\begin{itemize}
    \item $a_1$ is a weight for how heavily the generated assistance should rely on visuals.
    \item $a_2$ is a weight for the inclusion examples, such as videos or step-by-step guides, that should be included in the generated assistance.
    \item $a_3$ is a weight for the inclusion of practice problems.
    \item $a_4$ is a weight for how much guidance the provided assistance should give to players. For example, a low value here would mean that the player simply receives educational material and must learn, while a high value here would mean that the system gives significantly more detailed guidance on what the player should study.
    \item $a_5$ is a weight for how encouraging and emotionally-driven the dialogue given to the player should be.
\end{itemize}

All values are essentially treated as weights in the range $[0, 1]$ and interpreted by the personalized assistance generation module which might combine rule-based systems with large language models. For example, with the weights given above, a rule-based system could first interpret $a_1, a_2,$ and $a_3$, which are the weights for visuals, examples, and practice problems. We could then take an expert-created set of study materials and assign threshold values to each piece of content. Some images might only be shown to students who were assigned an $a_1$ value of $0.9$ or higher, with similar rules in place for example problems and practice problems. Meanwhile, $a_4$ and $a_5$ are more abstract, so a large language model could be used to dynamically regenerate existing educational content to add in more or less guidance based on $a_4$, and more or less encouragement based on $a_5$. With this simple example, students would see highly varied study material based on the RL system's decision-making, and student feedback on the usefulness of that material could then feed back into the RL to inform decisions for future students.

\section{Conclusions} \label{sec:con}
In conclusion, we present a general-purpose framework for personalized educational support in virtual educational systems. Integrating data collection, semantic structuring through ontologies, and multi-agent reinforcement learning, the proposed system allows for artificial intelligence to augment, rather than replace, human intelligence in the classroom. By applying the proposed system to the context of virtual education, researchers can more easily provide intelligent, personalized feedback and recommendations to students without the need for instructor intervention. With the modular nature of the proposed system, educational concepts can easily be adjusted and added, and reinforcement learning allows for the system to adapt to shifts in student preferences. Moving forward, we intend to focus on empirical validation of our system in an educational context, as well as further optimizations to the methodology, such as a deeper look at personalized assistance generation, and additional improvements to the reinforcement learning. Ultimately, the proposed system sets a unique and useful model for future researchers, advancing the field of personalized education by making it more accessible to implement.

\bibliographystyle{ieeetr}
\bibliography{refs.bib}

\begin{thebibliography}{10}

\bibitem{aiaugmentation}
K.-L.~A. Yau, H.~J. Lee, Y.-W. Chong, M.~H. Ling, A.~R. Syed, C.~Wu, and H.~G. Goh, ``Augmented intelligence: Surveys of literature and expert opinion to understand relations between human intelligence and artificial intelligence,'' {\em IEEE Access}, vol.~9, pp.~136744--136761, 2021.

\bibitem{grandchallenges}
NAE, ``Nae grand challenges: Advance personalized learning,'' 2009.

\bibitem{challengeofpersonalization}
A.~J. Bingham, J.~F. Pane, E.~D. Steiner, and L.~S. Hamilton, ``Ahead of the curve: Implementation challenges in personalized learning school models,'' {\em Educational Policy}, vol.~32, no.~3, pp.~454--489, 2018.

\bibitem{classsizes}
I.~Y. Johnson, ``Class size and student performance at a public research university: A cross-classified model,'' {\em Research in Higher Education}, vol.~51, no.~8, pp.~701--723, 2010.

\bibitem{onesizeissues}
S.~Goodman and H.~Bohanon, ``{A Framework for Supporting All Students: One-Size-Fits-All No Longer Works in Schools},'' {\em Loyola eCommons}, vol.~February, 2018.

\bibitem{aiineducation}
F.~AlShaikh and N.~Hewahi, ``Ai and machine learning techniques in the development of intelligent tutoring system: A review,'' in {\em 2021 International Conference on innovation and Intelligence for informatics, computing, and technologies (3ICT)}, pp.~403--410, IEEE, 2021.

\bibitem{aiineducation2}
C.-C. Lin, A.~Y.~Q. Huang, and O.~H.~T. Lu, ``{Artificial intelligence in intelligent tutoring systems toward sustainable education: a systematic review},'' {\em Smart Learn. Environ.}, vol.~10, pp.~1--22, Dec. 2023.

\bibitem{aiengagement}
C.~Vrabie, ``Education 3.0 -- ai and gamification tools for increasing student engagement and knowledge retention,'' in {\em Digital Transformation} (J.~Ma{\'{s}}lankowski, B.~Marcinkowski, and P.~Rupino~da Cunha, eds.), (Cham), pp.~74--87, Springer Nature Switzerland, 2023.

\bibitem{whatontology}
L.~Ding, P.~Kolari, Z.~Ding, and S.~Avancha, ``Using ontologies in the semantic web: A survey,'' {\em Ontologies: A Handbook of Principles, Concepts and Applications in Information Systems}, pp.~79--113, 2007.

\bibitem{ontomodeling}
H.~Yago, J.~Clemente, D.~Rodriguez, and P.~F. de~Cordoba, ``On-smmile: Ontology network-based student model for multiple learning environments,'' {\em Data \& Knowledge Engineering}, vol.~115, pp.~48--67, 2018.

\bibitem{ontomodellingits}
J.~Clemente, J.~Ramírez, and A.~{de Antonio}, ``A proposal for student modeling based on ontologies and diagnosis rules,'' {\em Expert Systems with Applications}, vol.~38, no.~7, pp.~8066--8078, 2011.

\bibitem{ontostandards}
S.~M. Rashid and D.~L. McGuinness, ``Creating and using an education standards ontology to improve education,'' in {\em Proceedings of the Workshop on Semantic Web for Social Good co-located with 17th International Semantic Web Conference, SW4SG@ISWC 2018, Monterey, California, USA, October 9, 2018., Monterey, California, USA, October 9, 2018} (K.~K. Waterman, ed.), vol.~2182 of {\em CEUR Workshop Proceedings}, CEUR-WS.org, 2018.

\bibitem{ontocontent}
E.~L. Baker, ``{CRESST Ontology-Based Educational Design: Seeing is Believing},'' {\em CRESST}, Dec. 2012.

\bibitem{reinforcementlearning}
C.~J. C.~H. Watkins, {\em Learning from Delayed Rewards}.
\newblock PhD thesis, King's College, Oxford, 1989.

\bibitem{rlconvergence}
Y.~Li, ``Deep reinforcement learning: An overview,'' 2018.

\bibitem{priormulti}
R.~Hare and Y.~Tang, ``Reinforcement learning with experience sharing for intelligent educational systems,'' {\em 2023 IEEE International Conference on Systems, Man, and Cybernetics}, 2023.

\bibitem{priortoe}
R.~Hare, Y.~Tang, and S.~Ferguson, ``(in press). an intelligent serious game for digital logic education to enhance student learning,'' {\em IEEE Transactions on Education}, 2024.

\bibitem{ddpg}
T.~P. Lillicrap, J.~J. Hunt, A.~Pritzel, N.~Heess, T.~Erez, Y.~Tassa, D.~Silver, and D.~Wierstra, ``Continuous control with deep reinforcement learning,'' 2019.

\bibitem{td3}
S.~Fujimoto, H.~van Hoof, and D.~Meger, ``Addressing function approximation error in actor-critic methods,'' 2018.

\bibitem{sac}
T.~Haarnoja, A.~Zhou, P.~Abbeel, and S.~Levine, ``Soft actor-critic: Off-policy maximum entropy deep reinforcement learning with a stochastic actor,'' 2018.

\bibitem{ppo}
J.~Schulman, F.~Wolski, P.~Dhariwal, A.~Radford, and O.~Klimov, ``Proximal policy optimization algorithms,'' 2017.

\bibitem{llmedu}
C.~K. Lo, ``What is the impact of chatgpt on education? a rapid review of the literature,'' {\em Education Sciences}, vol.~13, no.~4, p.~410, 2023.

\bibitem{aigcedu}
Z.~Bahroun, C.~Anane, V.~Ahmed, and A.~Zacca, ``Transforming education: A comprehensive review of generative artificial intelligence in educational settings through bibliometric and content analysis,'' {\em Sustainability}, vol.~15, no.~17, 2023.

\bibitem{rulebasededu}
N.~S. Raj and V.~Renumol, ``A rule-based approach for adaptive content recommendation in a personalized learning environment: An experimental analysis,'' in {\em 2019 IEEE tenth international conference on technology for education (T4E)}, pp.~138--141, IEEE, 2019.

\end{thebibliography}

\end{document}